\documentclass[12pt]{iopart}

\usepackage{graphicx,color} 
 
\begin{document}

\title{Crystal field splitting in  correlated systems with 
the negative charge transfer gap}

\author{A V Ushakov}
\address{II. Physikalisches Institut, Universit$\ddot a$t zu K$\ddot o$ln,
Z$\ddot u$lpicher Stra$\ss$e 77, D-50937 K$\ddot o$ln, Germany}
\ead{ushakov@ph2.uni-koeln.de}

\author{S V Streltsov}
\address{Institute of Metal Physics, S.Kovalevskoy St. 18, 620990 
Ekaterinburg, Russia}
\address{Ural Federal University, Mira St. 19, 620990 
Ekaterinburg, Russia}
\ead{streltsov@imp.uran.ru}

\author{D I  Khomskii}
\address{II. Physikalisches Institut, Universit$\ddot a$t zu K$\ddot o$ln,
Z$\ddot u$lpicher Stra$\ss$e 77, D-50937 K$\ddot o$ln, Germany}

\begin{abstract}
Special features of the crystal field splitting of $d-$levels 
in the transition metal compounds with the small or negative charge-transfer 
gap $\Delta_{CT}$ are considered. 
We show that in this case the 
Coulomb term and the covalent contribution to the
$t_{2g} - e_g$ splitting have different signs.
In order to check the theoretical predictions we carried out the ab-initio band structure 
calculations for Cs$_2$Au$_2$Cl$_6$, in which the charge-transfer gap is negative,
so that the $d-$electrons predominantly occupy low-lying bonding states. 
For these states the $e_g$-levels lie below $t_{2g}$ ones, which demonstrates that at least in 
this case the influence of the $p-d$ covalency on the total value of the crystal
field splitting is stronger than the Coulomb interaction 
(which would lead to the opposite level order). We also show that the states in conduction band are 
made predominantly of $p-$states of ligands (Cl), with small admixture of $d-$states of Au.

\end{abstract}

\maketitle

\section{Introduction}

One of the most important factors determining properties of the transition metal (TM) oxides is 
the splitting of $d$-levels when the TM ions are put in a crystal. 
 This crystal-field (CF) splitting largely determines magnetic and orbital state 
of respective ions, and thus their properties. In general there are two mechanisms, which have to be 
taken into account in the calculation of the CF splitting. 

One is the Coulomb interaction of $d$-electrons with charges of surrounding ions - mainly 
nearest-neighbor anions like O$^{2-}$ in oxides or F$^-$ and Cl$^{-}$ in fluorides and chlorides. 
Since the Coulomb repulsion with anions depends on the shape of the electron
density, the resulting corrections to the energy spectrum will be different for 
different $d-$orbitals.

The second mechanism which also leads to the CF splitting is the covalency, the hybridization between
$d-$states of the TM ions and ligand $p-$states. 
The $p-d$ hybridization (hopping integrals $t^{pd}$ in Eq.~(\ref{Ham}))
strongly depends on the symmetry of the local environment and will be very different for $d-$orbitals 
pointing to the ligands ($e_g$ orbitals in the case of a octahedral
surrounding) or looking in between of them ($t_{2g}$ orbitals in the octahedra),
which would lead to their splitting. Here and below we will discuss the 
octahedral symmetry. 

We show below that the role of these
two mechanisms may strongly differ in conventional TM compounds and in the materials with
the negative charge-transfer (CT) gaps~\cite{ZSA,Kh}, with a (e.g. oxygen) holes in the ground state.

\section{The basic model}

The standard model describing transition metal (TM) ions in a crystal is
\begin{eqnarray}
\nonumber
H &=& \epsilon_{d0} \sum_{i \alpha } d_{i\alpha\sigma}^{\dag} d_{i\alpha\sigma} 
+ \epsilon_{p} \sum_{j \beta} p_{j\beta\sigma}^{\dag} p_{j\beta\sigma} \\
\nonumber
&+& \sum_j t^{pd}_{ij \alpha \beta} (d_{\alpha\sigma}^{\dag}p_{j\beta\sigma} + h.c.) \\
&+& \sum_{(\alpha \sigma) \neq (\beta \sigma')} U^{dd}_{\alpha \beta} n_{i\alpha\sigma}
n_{i\beta\sigma'}.
\label{Ham}
\end{eqnarray}

Here $d_{i\alpha\sigma}^{i\dag}$, $d_{i\alpha\sigma}$ 
are the creation and annihilation operators of 
$d-$electrons at a site i in the orbital state $\alpha$ with spin $\sigma$, 
$p_{j\beta\sigma}^{\dag}$, $p_{j\beta\sigma}$ describe different $p-$electrons at the $j-$th
ligand surrounding given TM site. 
The system is also characterized by the on-site Coulomb (Hubbard) interaction $U^{dd} = U$, and by 
the charge-transfer energy $\Delta_{CT}$, defined as the energy of the transition
\begin{eqnarray}
\label{CT-reaction}
d^n p^6 \to d^{n+1} p^5,
\end{eqnarray}
which in the electron representation used in the Hamiltonian~(\ref{Ham}) is
\begin{eqnarray}
\label{CT-deter}
\Delta_{CT} = \epsilon_{d0} + nU - \epsilon_p = \epsilon_d - \epsilon_p.
\end{eqnarray}
Here we introduced the renormalized energy of $d-$levels 
\begin{eqnarray}
\label{d-energy}
\epsilon_d = \epsilon_{d0} + nU                                                   
\end{eqnarray}
- the effective $d-$energy levels to which an electron is transferred 
from $p-$states of a ligand in the ``reaction''~(\ref{CT-reaction}).~\cite{FT1} 
Further on in this paper by $d-$levels shown e.g. in 
Figs.~\ref{posit-delta} and \ref{neg-delta} we have in mind these 
renormalized $d-$levels~(\ref{d-energy}).

In principle one should include in this Hamiltonian some other terms, such as the 
intra-atomic Hund's rule exchange $J_H$, Coulomb repulsion for ligand $p-$electrons 
$U_{pp}$, etc. For our purposes, however, the simplified model~(\ref{Ham}) is sufficient.

\section{Positive charge-transfer energy, $\Delta_{CT}>0$}
\begin{figure}[t]
  \centering
  \includegraphics[clip=true, width=0.5\textwidth]{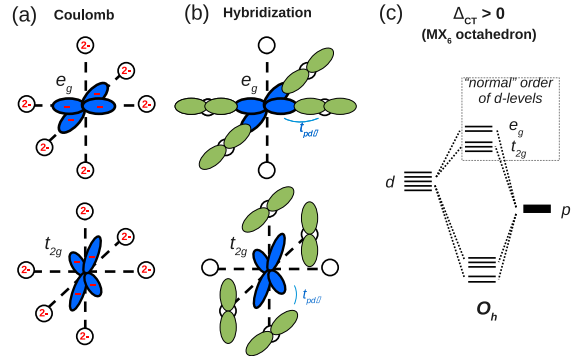}
  \caption{(color online)
  Charge-transfer gap $\Delta_{CT}>0$. There are two contributions to the
  resulting CF splitting caused by (a) the Coulomb
  interaction and (b) the $p-d$ hybridization (or
  covalency). The $e_g-$orbitals 
  (the $x^2-y^2$ is shown as an example) 
  pointing to the ligands have larger hybridization and stronger repulsion
  from them and as 
  a result lie above the $t_{2g}-$levels (the $yx-$orbital is sketched) . The ligands forming octahedra 
are shown as open circles, 
  $d-$orbitals are painted in blue and ligand $p-$orbitals in green. For
  the Coulomb contribution the charges (not the signs of the wave-functions) 
  are indicated. The levels with predominantly $d-$character are marked by the dashed ``box''. 
   }
  \label{posit-delta}
\end{figure}

In a most typical situation for the TM compounds, when the ligand $p-$levels lie deep below $d-$levels 
of the TM ions, both the Coulomb and the hybridization contributions to the CF splitting lead to the same 
sequence of $d-$levels. This case is
characterized by a large positive CT gap $\Delta_{CT}$, defined in Eq.~(\ref{CT-deter}). 
As explained above and
as is clearly seen in Fig.~\ref{posit-delta}(a), the Coulomb repulsion of $d$-electrons 
with negatively-charged ligands is stronger for $e_g-$electrons with the lobes 
of the electron density directed towards the ligands than for 
$t_{2g}-$orbitals pointing in between the anions.  This leads to the CF splitting shown in 
Fig.~\ref{posit-delta}(c) such that $e_g-$orbitals lie {\it above} $t_{2g}$. This 
sequence of $d-$levels is referred as ``normal''.

The same ``normal'' sequence of $d-$levels is also caused by the $p-d$ hybridization, i.e. by covalent 
contribution. Since the hopping matrix elements 
$e_g-p$ ($t_{pd\sigma}$) are larger than hoppings between $t_{2g}-$orbitals
and $p-$orbitals of ligands ($t_{pd\pi}$), the $e_g-$levels are pushed
up stronger than $t_{2g}$,
see Figs.~\ref{posit-delta}(b), (c).~\cite{Harrison}

Relative importance of these two contributions to the total value of the CF splitting
depends on the compound, but in the usual situation of $\Delta_{CT}>0$
they at least work in the same direction. 

In the early stage of the development of the CF theory
it was argued that an account of the real shape of ligand orbitals may
change the sign of the Coulomb contribution, obtained in the point-charge 
model.~\cite{Kleiner-52} However, it was shown later that in order to consider ionic term 
in a correct way the $d-$electron wave functions must be orthogonalized to the ligands.~
\cite{Tanabe-56}
As a result, the total ionic contribution to the CF splitting has the
same sign as the covalent term in the case of  $\Delta_{CT}>0$.~\cite{Sugano-63}

Note that for the {\it bonding} levels (the lower levels in Fig.~\ref{posit-delta}(c)) the covalency 
contribution 
gives the sequence of levels opposite to the ``normal'': the levels with the $e_g-$symmetry have 
stronger bonding-antibonding splitting and lie {\it below} those with the $t_{2g}-$symmetry. 
But for large positive CT gap these states are predominantly made of the combination 
of $p-$states of ligands with a proper symmetry, with only a small admixture of $d-$states, 
and one rarely discuss these states, which lie deep below the Fermi energy. All interesting 
phenomena take place in the partially-filled antibonding states, which have predominantly 
$d-$character and which are responsible e.g. for magnetism, orbital ordering etc. 
But this changes drastically in case of the negative CT gap $\Delta_{CT}<0$, which we now consider.

\section{Negative charge-transfer energy, $\Delta_{CT}<0$}
\begin{figure}[t]
  \centering
  \includegraphics[clip=true, width=0.47\textwidth]{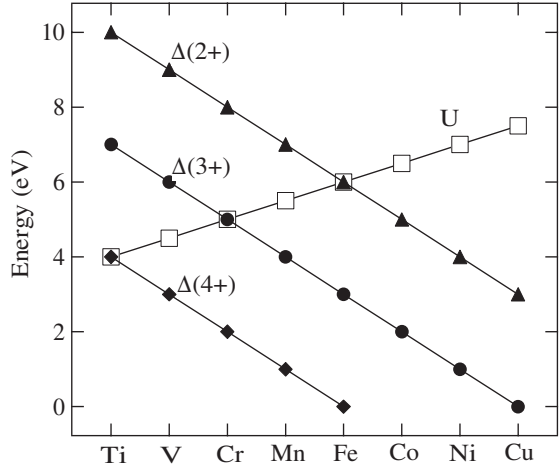}
  \caption{The charge-transfer (CT)
gap $\Delta_{CT}$ and the on-site 
Coulomb repulsion $U$ values for different $3d$-ions, after T. Mizokawa.~\cite{Mizokawa}}
  \label{DeltaCT}
\end{figure}

The situation is quite different for the systems with small and especially negative 
CT gaps. In this case the Coulomb and covalent contributions to the splitting of 
$d-${\it levels} have different signs. 
The negative CT gap is met in the compounds based on the late TM ions with unusually 
high valence, or high oxidation state~\cite{Kh}, 
like Fe$^{4+}$ or Cu$^{3+}$, see Fig.~\ref{DeltaCT}, compiled by Mizokawa on the basis of 
the data of~\cite{Bocquet-92,Mizokawa}. 

The negative CT energy means
that it is favorable to transfer an electron from the ligand $p$-shell to a $d$-level,
i.e. to perform the ``reaction''~(\ref{CT-reaction})
already in the ground state. Such a process is sometimes called self-doping.~\cite{Khomskii-97, Korotin-97}
On the language of energy levels a regime with the negative 
CT energy corresponds to the situation when the initial  {\it renormalized}
$d-$levels $\epsilon_d$ in~(\ref{d-energy}) lie below the ligand $p-$levels.

\begin{figure}[t]
  \centering
  \includegraphics[clip=true, width=0.5\textwidth]{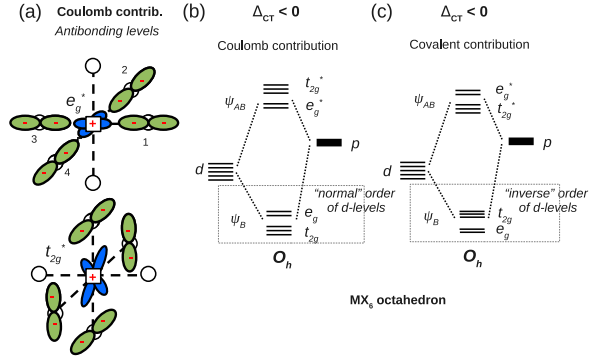}
  \caption{
  (color online)
  Charge-transfer (CT) gap $\Delta_{CT}<0$. The crystal-field splitting
  caused by the Coulomb interaction (b) and due to the covalency (c) are shown.
  Examples of the antibonding orbitals are sketched in the panel (a).
  Ligands forming octahedra are shown as open circles, 
  $d-$orbitals are painted in blue and ligand $p-$orbitals in green. For
  the Coulomb contribution the charges (not the signs of the wave-functions) 
  are indicated. The levels with predominantly $d-$character are marked by the dashed ``box''.
  }
  \label{neg-delta}
\end{figure}

We start with an analysis of the covalent contribution to the CF splitting,
Fig.~\ref{neg-delta}(c). 
In contrast to the case of $\Delta_{CT} >0$ it is now the lower bonding orbitals 
that have predominantly $d-$character with a small admixture of $p-$states
\begin{equation}
\label{eq:bond}
\Psi_{B} = \alpha \vert d \rangle + \beta \vert p \rangle, \qquad
\alpha^{2} + \beta^{2} = 1, \qquad \alpha \gg \beta.
\end{equation}

For these bonding states the covalency leads to the splitting such that 
the $t_{2g}-$levels lie above $e_g-$ones, i.e. this covalency contribution leads to an 
``inverted'' CF splitting for $d-$electrons.

These bonding states are actually completely filled. 
The electrons of partially-filled levels 
will actually be in the antibonding orbitals $t_{2g}^{\ast},e_g^{\ast}$. 
Due to a stronger ${pd\sigma}$ hybridization of orbitals having $e_g-$symmetry with the
ligand $p-$shell, the bonding-antibonding splitting of $e_g-$orbitals is larger than that of 
$t_{2g}-$orbitals. As a result the order of antibonding levels is ``normal'', the same as in 
Fig.~\ref{posit-delta} ($t_{2g}^{\ast}-$levels are below $e_g^{
the \ast}-$subshell). But, in 
contrast to the case of a positive CT gap, here the wave function of 
the antibonding states will be 
\begin{equation}
\label{eq:antibond}
\Psi_{AB} = \alpha^{\ast} \vert d \rangle + \beta^{\ast} \vert p \rangle, \qquad (\alpha^{\ast})^{2} + (\beta^{\ast})^{2} = 1, 
\qquad \alpha^{\ast} \ll \beta^{\ast},
\end{equation}
i.e. these states have predominantly $p-$character. In effect the resulting state would be the state 
with the self-doping described above, i.e. with a large fraction of oxygen holes.


The Coulomb contribution, however, would lead to the opposite sequence of $d-$levels,
compared to the covalent term. 
For the $d-$levels themselves it would have the same sign as for the case of 
$\Delta_{CT}>0$, and due to a stronger repulsion 
from ligands the $e_g-$levels (bonding) would be pushed {\it above}
$t_{2g}$, opposite to the influence of the hybridization.

One has to somewhat specify these arguments. In the case of $\Delta_{CT}<0$
the ground state should be better taken not as $d^np^6$, but rather as $d^{n+1}p^5$,
according to~(\ref{CT-reaction}). This will change the magnitude of the Coulomb
terms, the ting effect: still $e_g-$electrons would repel
stronger from the ligands (with smaller charge, e.g. O$^{1.5-}$ instead of
O$^{2-}$). Thus in the case of the negative CT gap the contributions of the 
Coulomb interaction and of the $p-d$ covalency to the CF splitting of 
$d-$states is opposite. 

The same can be said about the antibonding levels.
As follows from Eq.~(\ref{eq:antibond}), the antibonding levels 
have predominantly $p-$character, but the combinations of ligand $p-$electrons with 
$e_g$-symmetry shown in Fig.~\ref{neg-delta}(a), (e.g. the $\sigma$ combination 
$p_{\sigma} = \frac{1}{\sqrt{4}}[p_{1x}-p_{2y}-p_{3x}+p_{4y}]$, which hybridizes 
with $d_{x^2-y^2}$-orbital) would have stronger attraction to a positively-charged $d-$ion 
than the $t_{2g}^{\ast}-$orbitals, made out of the $\pi-$combination of ligand orbitals.
As a result the antibonding $t_{2g}^{\ast}-$levels due to this factor would lie higher than 
$e_g^{\ast}$ (Fig.~\ref{neg-delta}(b)). 

Thus we see that, in contrast to the ``normal'' case of positive CT gap, where both contributions 
to the CF splitting, the Coulomb contribution and the covalency, lead to 
qualitatively the same order of levels (antibonding $e_g$-levels are above $t_{2g}$, and 
vice versa for bonding states), for negative CT gaps the situation is different: these 
two contributions lead to the {\it opposite} level order. Therefore by 
studying the actual order of these levels, experimentally or theoretically, one can 
get some information about which of these contributions is in fact stronger. 
We undertook such a study on the example of a system with the negative CT gap - Cs$_2$Au$_2$Cl$_6$.

\section{Crystal field in Cs$_2$Au$_2$Cl$_6$}
\begin{figure}[t]
  \centering
  \includegraphics[clip=true, width=0.4\textwidth]{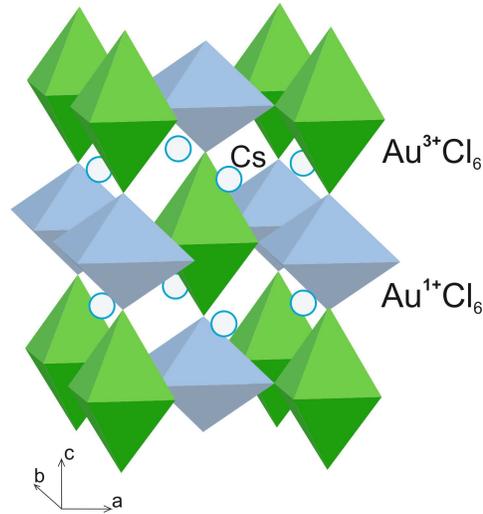}
  \caption{The crystal structure of Cs$_2$Au$_2$Cl$_6$. }
  \label{crystal-structure}
\end{figure}

Cs$_2$Au$_2$Cl$_6$ should formally contain the Au$^{2+}(d^9)$ ions. However, it is 
well-known in chemistry 
that this valence state of gold is unstable and is practically never observed 
experimentally: Au is known to exist in a valence Au$^{1+}$ or Au$^{3+}$. 
This also happens in this material: there occurs a spontaneous charge 
disproportionation (CD) into Au$^{1+}$ (Au2) and Au$^{3+}$ (Au1), so that there appears 
in this compound two inequivalent Au positions, with checkerboard ordering 
in a cubic perovskite lattice.~\cite{Elliot-38,Verschoor-74}
, see Fig.~\ref{crystal-structure}.
 In effect even 
the formula of this material is usually written 
not as CsAuCl$_3$, as for usual perovskites, but as
Cs$_2$Au$_2$Cl$_6$ = Cs$_2$Au$^{1+}$Au$^{3+}$Cl$_6$.

We carried out the ab-initio band structure calculations of this 
material, have shown that it has negative CT gap, and analyzed the CF
splitting.

The ab-initio band-structure calculations of Cs$_2$Au$_2$Cl$_6$ were performed 
within the frameworks of the Density Functional Theory (DFT) and the Generalized Gradient 
Approximation (GGA). We used the PW-SCF code with the ultrasoft version
of pseudopotentials~\cite{Vanderbilt-90,PW-SCF}.
A maximum cut-off energy for the plane waves was chosen to be 200 eV. 
The Brillouin-zone (BZ) integration in the course of self-consistency
interactions was performed over a mesh of 72 k-points.
The structural data for ambient pressure were taken from Ref.~\cite{Denner-79}.
The Wannier functions were constructed using the formalism described 
in Ref.~\cite{Korotin-08}.

\begin{figure}[hbp!]
  \centering
  \includegraphics[clip=true, width=0.4\textwidth]{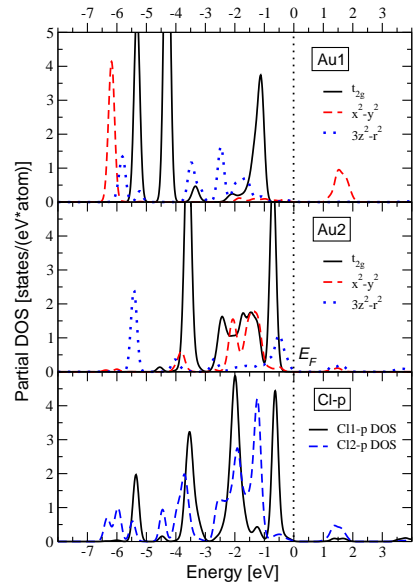}
  \caption{Partial density of states 
for  Cs$_2$Au$_2$Cl$_6$ at ambient 
pressure. Au1 represent atom, which should have oxidation
state 3+, while Au2 - 1+ (according to valence bond sum
analysis of the crystal structure). The Cl$-p$ partial density of states  is
the normalized per one Cl for two classes of ``apical'' Cl1 and ``plane'' Cl2 in the octahedra.
The Fermi level is taken as zero.}
  \label{PDOS}
\end{figure}

The partial densities of states (PDOS) are presented in Fig.~\ref{PDOS}.
One may clearly see that indeed the lowest states in the valence band have
largest contributions from the $Au-d$, not $Cl-p$ states, so that
the CT energy is indeed negative in this compound. 
The states closest to the top of the valence band, as well as the states in 
the conduction band, are mainly of $p$-character.

The total occupation of Au2 $5d-$shell is 9.6, which is close to the $d^{10}$ 
configuration, expected for Au$^{1+}$. However the number of $d$-electrons for Au1 
is 9.3, much larger than 8 electrons expected from the naive ionic picture 
for the $3+$ oxidation state of Au. Thus one should represent this states not as
Au$^{3+}$ ($d^8$), but rather as $d^9 \underline{L}$ or even $d^{10} \underline{L^2}$, where $\underline{L}$ denotes
ligand hole. This is due to the self-doping effect 
described above. Au$^{3+}$ ions turned out to be Jahn-Teller active, but non-magnetic,
with the configuration $t_{2g}^{6} e_{g \uparrow}^1 e_{g \downarrow}^1$
(or rather $d^{10} \underline{L} \uparrow  \underline{L} \downarrow$).

As it was shown in previous two sections, a larger hybridization leads to a larger 
bonding-antibonding splitting, independent on the sign of the $\Delta_{CT}$.
The Cl octahedron surrounding Au1 is elongated along $z-$axis. 
Then the bonding-antibonding splitting for the orbital of $x^2-y^2$-symmetry should be the largest among 
all of Au $5d-$orbitals. This is exactly what we see in the upper panel of 
Fig.~\ref{PDOS}. The Coulomb contribution to the crystal-field splitting
would stabilize just the opposite sequence of levels, see Fig.~\ref{neg-delta}(b). 
This means that at least in this system the Coulomb contribution
to the total value of the CF splitting is much less than the 
covalent term. 

\begin{figure}[t]
  \centering
  \includegraphics[clip=true, width=0.4\textwidth]{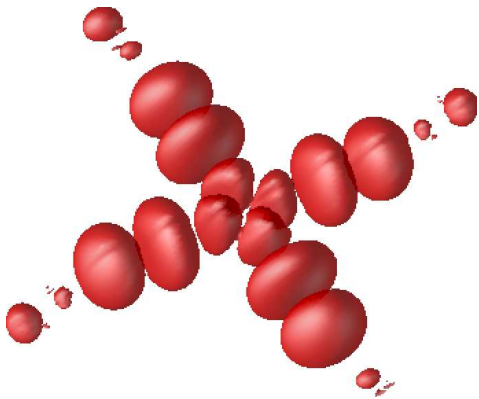}
  \caption{The charge-density corresponding to the antibonding orbital of $x^2-y^2-$symmetry centered on 
Au1 site, nominally $Au^{3+}$. The central part corresponds to the contribution of the
real Au$-5d$ $x^2-y^2$ orbital, while the rest
is a $\sigma$ combination of the Cl$-p$ orbitals.}
  \label{x2y2-orbital}
\end{figure}

It is important to mention that the conduction band
is basically formed by the $\sigma$ combination of Cl$-p$ states with
$x^2-y^2$ symmetry.  These are actually 
the antibonding $e_g^*$ orbitals of the type (6), shown in Fig.~\ref{neg-delta}(a). The charge
density for this orbital, obtained using the Wannier function projection technique,
is shown in Fig.~\ref{x2y2-orbital}. The detailed calculations
show that about $\sim$20 \% of the charge-density corresponding to this
orbital has a $5d-$character, while the main part $\sim 80$ \% comes 
from the Cl$-p$ orbitals.

A similar analysis can be performed for the compressed along the $z$-axis Au2O$_6$ octahedra. Here again the covalent contribution
dominates and pushes down the bonding orbital of $3z^2-r^2$-symmetry, which
has the largest hybridization with the Cl$-p$ states.

The question which naturally arises is how general is the conclusion reached above, 
that the covalency contribution to the CF splitting is (much) larger than the 
Coulomb contribution. Generally speaking, one could think that the covalency 
contribution to the $t_{2g}-e_g$ splitting, which is
\begin{equation}
\Delta_{CF}^{cov.} \sim \frac{t_{pd\sigma}^2-t_{pd\pi}^2}{\Delta_{CT}},
\end{equation}
 could become small for very large $\vert \Delta_{CT} \vert$. It is not surprising that 
this contribution dominates for systems with small or slightly negative CT gap 
(apparently Cs$_2$Au$_2$Cl$_6$ belongs to this class). Different 
approaches~\cite{Sugano-63, Haverkort-11}, however, demonstrate that it is also the 
case for the values of parameters typical for ``normal'' TM compounds, such as 
KNiF$_3$~\cite{Sugano-63} 
or 
NiO.~\cite{Haverkort-11} Nevertheless, one can not exclude that in 
some cases (e.g. for the early 3d metals) the situation could be different and 
the Coulomb contribution would become comparable or larger than the covalency one.

\section{Summary}

In this paper we analyzed possible features of the crystal field splitting in 
different situations, and argued that for the negative charge-transfer gaps, in contrast 
to the ``normal'' situation with $\Delta_{CT}>0$, different contributions to the 
crystal field splitting, the Coulomb contribution and that of the covalency with the 
ligands, act in opposite direction: the Coulomb contribution always pushes the $t_{2g}$ levels 
below the $e_g$ ones, whereas the $p-d$ hybridization would lead to the opposite order of the 
bonding levels (which for negative charge transfer gap have predominantly $d-$character). 
The study of a concrete example of Cs$_2$Au$_2$Cl$_6$ demonstrates that this system  
indeed  has negative charge-transfer gap, so that the states in the conduction band 
are largely made of $p-$states of Cl. The bonding states of $d-$character lie $\sim$2-6 eV 
below the Fermi energy, and in this region the $e_g$ levels lie below $t_{2g}$ ones. 
Thus in this particular system the $p-d$ hybridization dominates and determines the 
sequence of levels. We discuss in conclusion whether the opposite situation could be 
possible, in which case the sequence of $d-$levels would be determined by the 
Coulomb contribution and might be different from the usual one.

\section{Acknowledgments}

We acknowledge fruitful communications with I. Mazin and M. Haverkort. We are very grateful to D. Korotin,
for the code used for the projection on the Wannier functions, and especially to
T. Mizokawa for providing us and allowing to use the Fig.~\ref{DeltaCT}.

This work was supported by the German projects SFB 608, DFG GR 1484/2-1 and FOR 1346, by the European project SOPRANO, 
by Russian projects 
RFBR 10-02-00046 and 10-02-96011,
by the program of President of Russian Federation MK-309.2009.2,
the Russian Federal Agency of Science and Innovation N 02.740.11.0217 and
the scientific program ``Development of scientific potential of universities'' 
N 2.1.1/779.


\section*{References}
\begin{thebibliography}{50}
\bibitem{ZSA} Zaanen J, Sawatzky G~A  and Allen J~W 1985 {\it Phys. Rev. Lett.} {\bf 55} 418 
\bibitem{Kh} Khomskii D I 1997 {\it Lithuanian Journal of Physics} {\bf 37}, 2001 {\it cond. mat.} 0101164
\bibitem{FT1} Often, e.g. in cuprates, one works  in the hole representation, 
and then the energy of a transition~(\ref{CT-reaction}) 
in this 
representation would be simply
$\Delta_{CT} = \tilde \epsilon_{p-hole} - \tilde \epsilon_{d-hole})$. 
\bibitem{Harrison} Harrison  W A  1999 {\it Elementary Electronic Structure} (World Scientific, Singapore)
\bibitem{Kleiner-52} Kleiner  W H 1952 {\it J. Chem. Phys.} {\bf 20} 1784
\bibitem{Tanabe-56} Tanabe Y and Sugano S 1956 {\it J. Phys. Soc. Japan} {\bf 11} 864
\bibitem{Sugano-63} Sugano S and Shulman R G 1963 {\it Phys. Rev.} {\bf 130} 517 
\bibitem{Bocquet-92} Bocquet A E, Mizokawa T, Saitoh T, Namatame H, and  A. Fujimori A 1992 {\it Phys. Rev. B} {\bf 46} 3771
\bibitem{Mizokawa} Mizokawa T {\it Priv. Commun}
\bibitem{Korotin-97} Korotin M A, Anisimov V I, Khomskii D I and Sawatzky G A 1997 {\it Phys. Rev. Lett.} {\bf 80} 4305 
\bibitem{Khomskii-97} Khomskii D I and Sawatzky G A 1997 {\it Solid State Commun.} {\bf 87} 102
\bibitem{Elliot-38} Elliott N and Pauling S L 1938 {\it J. Am. Chem. Soc.} {\bf 60} 1846
\bibitem{Verschoor-74} Tindermanns J C M v. Eijndhoven and Verschoor S G C 1974 {\it Mater. Res. Bull.} {\bf 9} 1667 
\bibitem{Vanderbilt-90} Vanderbilt D 1990 {\it Phys. Rev. B.} {\bf 41} 7892  
\bibitem{PW-SCF} Baroni S \textit{et al.,} PWscf
(Plane-Wave Self-Consistent Field) codes, http://www.pwscf.org
\bibitem{Denner-79} Denner W, Schulz H, and  D'Amour H 1979 {\it Acta Crystallogr., Sect. A: Cryst. Phys., Diffr., 
Theor. Gen.  Crystallogr} {\bf 35} 360 
\bibitem{Korotin-08} Korotin Dm, Kozhevnikov A V, Skornyakov S L, Leonov I,
 Bingelli N, Anisimov V I and Trimarchi G 2008 {\it Eur. Phys. J.} {\bf B 65} 91  
\bibitem{Haverkort-11} Haverkort M et al. {\it to be published}



\end{thebibliography}
\end{document}